\documentclass[prl,twocolumn,superscriptaddress]{revtex4}
\usepackage{amsmath,amssymb,graphicx,xcolor,bm,soul}

\makeindex

\begin{document}

\title{Solving NP-hard problems with bistable polaritonic networks}

\author{O. Kyriienko}
\affiliation{NORDITA, KTH Royal Institute of Technology and Stockholm University, Roslagstullsbacken 23, SE-106 91 Stockholm, Sweden}

\author{H. Sigurdsson}
\affiliation{Science Institute, University of Iceland, Dunhagi-3, IS-107 Reykjavik, Iceland}

\author{T. C. H. Liew}
\affiliation{Division of Physics and Applied Physics, School of Physical and Mathematical Sciences, Nanyang Technological University, 21 Nanyang Link, Singapore 637371}

\begin{abstract}
A lattice of locally bistable driven-dissipative cavity polaritons is found theoretically to effectively simulate the Ising model, also enabling an effective transverse field. We benchmark the system performance for spin glass problems, and study the scaling of the ground state energy deviation and success probability as a function of system size. As particular examples we consider $\mathrm{NP}$-hard problems embedded in the Ising model, namely graph partitioning and the knapsack problem. We find that locally bistable polariton networks act as classical simulators for solving optimization problems, which can potentially present an improvement within the exponential complexity class.
\end{abstract}

\date{\today}

\maketitle

\textit{Introduction.} Solving complex optimization problems is highly demanded in various fields of science and information technologies, ranging from economics~\cite{Bouchaud_JSP2013} and finances~\cite{Durlauf_PNAS1999} to biology and physics~\cite{bryngelson_spin_1987, Pierce2002, nishimori_statistical_2001}. While certain problems can be solved deterministically in polynomial time and belong to class $\mathrm{P}$, many optimization problems do not have a deterministic solution. Namely, these correspond to tasks where the number of possible solutions scales exponentially (non-polynomially, thus $\mathrm{NP}$), and the true optimum should be searched for probabilistically. Methods include simulated annealing~\cite{Dowsland2012}, Monte-Carlo sampling \cite{Swendsen1986}, ant colony optimization \cite{Dorigo1996}, and genetic protocols~\cite{McCall2005}, which represent improvement on brute force (greedy) algorithms. The examples of $\mathrm{NP}$ problems are satisfiability, graph partitioning, and Hamiltonian cycles problems (e.g. travelling salesman), among others \cite{Lucas_FroPhy2014}. They include $\mathrm{NP}$-hard problems, which correspond to a search for the exact value of the optimal solution.

As $\mathrm{NP}$-hard problems are ubiquitous in nature and their efficient solving strategies represent a major milestone in many areas, the problem has attracted much attention in computational science. One of the possible approaches was suggested in the field of quantum computing, where a quantum adiabatic algorithm searches for the ground state of an associated spin glass-type Hamiltonian~\cite{Farhi_Science2001,Kadowaki_PRE1998}, which can be mapped to the solution of NP-complete problems \cite{Barahona1982, Lucas_FroPhy2014}. The strategy also serves as a goal for large scale quantum annealers built by DWave Inc.~\cite{Dickson_Nature2013}, though operating in the open system regime. However, thus far there is no significant evidence to suggest that there is an efficient (i.e. polynomial) quantum algorithm to solve $\mathrm{NP}$ tasks. Up to date, only constant speed-up was demonstrated~\cite{Denchev2016}. This poses the question of whether alternative strategies using classical analog simulating devices can provide similar advantages. Recently, one of these devices---a degenerate optical parametric network---attracted attention as a possible Ising model solver~\cite{Marandi_NatPhys2014, Inagaki_Science2016, McMahon_Science2016, Clements_PRA2017}, and has shown scalability potential. Other considered options include pure~\cite{AspuruGuzik_NatPhys2012} and hybrid photonic~\cite{Angelakis_Springer2017} quantum simulators.

A new emerging platform for classical simulation of effective spin models is the nonlinear system of exciton-polariton lattices~\cite{Lai_Nature2007, Kim_NatPhys2011, Tosi_NatCom2012, Jacqmin_PRL2014, Winkler_NJP2015, Berloff2016, Klembt_APL2017, Whittaker_PRL2018, Hamid2018, Suchomel2018}. Recently, several experiments reported the generation of real space lattices of polaritonic wells, where each node corresponds to a coherent non-equilibrium condensate of polaritons~\cite{Carusotto2013,Byrnes2014}. Coupling was realized through the delocalized photonic component, and the exciton-exciton interaction provides nonlinearity. So far classical simulators for XY-type models were considered \cite{Berloff2016, Berloff2018_2,Kalinin_2018ArXiv}, as well as spin chains~\cite{Ohadi_PRL2017}.

In this Letter we provide a general method to encode classical spin in the nonlinear polaritonic system, and show its ability to find the ground state configuration for the effective Ising model. This is based on the mapping of high/low intensity polaritonic states into a binary information. The approach is inherently nonlinear, and is largely different from previously demonstrated phase-encoded Ising \cite{McMahon_Science2016, Clements_PRA2017} and XY \cite{Berloff2016, Berloff2018_2} simulators, both in driven-dissipative and coherent circuit settings \cite{Goto2017}. We propose a feedback scheme that provides all-to-all coupling, and find that the suggested encoding allows to tackle concrete optimization problems without extra overhead required by mapping from other models with sparse connectivity or bias-free systems \cite{Cubitt2016}. This for instance can lead to huge improvements for system size, as an all-to-all connected Ising model with $O(100)$ spins would require up to $O(10^6)$ auxiliary spins to be simulated with a 2D nearest neighbor connected graph \cite{Cubitt2016}.

We consider actual optimization problems and show that native Ising encoding is beneficial for small scale optimizers. As an example, we apply bistable polaritonic networks to the NP-hard graph partitioning problem, where Ising-type interaction is encoded into polaritonic intermode tunneling. Then, we introduce bias in the system and apply the system to the knapsack problem, widely used in economics~\cite{KellererBook,Deniz2001}. Finally, we discuss possible implementation using existing technology.

\textit{Classical spin models with bistable states.} To begin, we consider the driven-dissipative nonlinear Schr\"odinger equation describing the evolution of a system of spatially separate but interconnected nonlinear optical resonators, each containing a complex field amplitude $\psi_n$:
\begin{equation}
i\frac{\partial\psi_n}{\partial t}=\left(-\Delta_n(t)-\frac{i}{2}+|\psi_n|^2\right)\psi_n+F_n(t)+\sum_mJ_{nm}\psi_m.
\label{eq:GP}
\end{equation}
$F_n$ represents the amplitude of a coherent driving field acting on the mode $n$. We work in the frame oscillating at the frequency of this driving field, which we assume to be the same for all resonators. $\Delta_n$ represents the detuning between the driving frequency and the resonant frequency of the mode $n$. We allow for both $F_n$ and $\Delta_n$ to be slowly varied over time. The dissipation in the system is represented by the term $-i/2$, where, without loss of generality, we have taken $t$ to have units of the inverse dissipation rate in the system. We account for a repulsive (self-defocusing) Kerr nonlinearity, the strength of which is scaled into the definition of $\psi_n$, again without loss of generality. The term $J_{nm}$ allows for coherent coupling between different modes, and in general enables all-to-all connectivity. While we identify exciton-polaritons in micropillar cavities as a potential implementation of the presented model, it also applies to a range of other nonlinear driven-dissipative bosonic systems (e.g., photonic crystals, superconducting circuits, driven-dissipative superfluids \cite{Labouvie2016}, etc.). Under coherent excitation Josephson coupling is the most accessible \cite{Rodriguez2016}, but we expect that similar results could be obtained with dissipative coupling mechanisms (i.e., imaginary $J_{nm}$).

\begin{figure}[t!]
\includegraphics[width=\columnwidth]{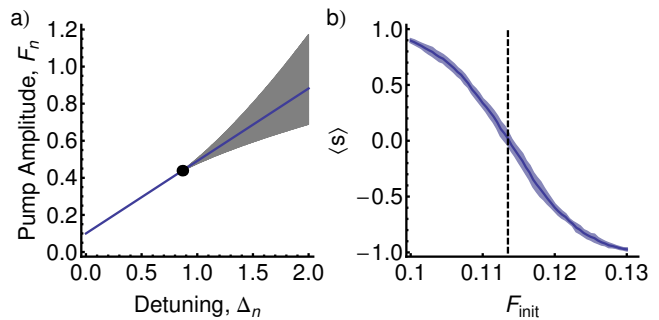}
\caption{a) Bifurcation diagram of a single nonlinear driven-dissipative system, described by Eq.~\eqref{eq:GP}. In the shaded region the system is bistable, while in the unshaded region it is monostable. The solid point marks the bifurcation point and the solid line a possible ramping of system parameters. b) Dependence of the average effective spin population on the initial pump amplitude after crossing the bifurcation point, with $\Delta_n(t)$ and $F_n(t)$ increasing linearly as stated in the text. The vertical line marks the value at which the effective spin population is zero on average, corresponding to $F_\mathrm{init}^{0}=0.1135$, and the noise level is set to $\theta_0=0.4$. The shading around the curve in b) indicates the statistical error.}
\label{fig:bifurcation}
\end{figure}

It is instructive to first consider a single isolated mode. Taking $\Delta_n$ and $F_n$ to be constant, and setting the time derivative in Eq.~\eqref{eq:GP} to zero yields a cubic equation for the stationary state intensity $|\psi_n|^2$, namely,
$\left(\left(\Delta-|\psi_n|^2\right)^2 + 1/4\right)|\psi_n|^2=|F_n|^2.$
In general the cubic equation yields three solutions; whether or not they are real depends on parameters. The possibilities are illustrated in Fig.~\ref{fig:bifurcation}a, which shows two types of behavior: in the unshaded region there is only one real stationary solution corresponding to monostable behavior, while in the shaded region all three solutions are real. In this latter case two of the solutions are stable and have different intensities, while the third solution is unstable. Consequently the system is considered bistable~\cite{Baas2004}, with both low or high intensity states possible under the same conditions. In Fig.~\ref{fig:bifurcation}a we can also identify a bifurcation point separating the bistable and monostable regions, at the critical pump amplitude $F_c=3^{-3/4}$ and critical detuning $\Delta_c=\sqrt{3}/2$. The solid line in Fig.~\ref{fig:bifurcation}a represents a possible slow (adiabatic) ramping of the parameters $F_n(t)$ and $\Delta_n(t)$ according to $\Delta_n(t)=\Delta_c t/\tau$ and $F_n(t)=F_{n,\mathrm{init}} + \left(F_c-F_{n,\mathrm{init}}\right)t/\tau$. Here, $F_{n,\mathrm{init}}$ defines the initial value of $F_n$ and $\tau$ is the time at which the bifurcation point is reached. Following this path of parameters, a dissipative phase transition is expected at the bifurcation point~\cite{FossFeig2017}, where the system must choose one of the two possible bistable states to lie in thereafter. The interplay of local bistability and Josephson coupling has been studied previously in Kerr nonlinear lattices, showing lattice solitons~\cite{Yulin2008,Egorov2013,Kartashov2016}, various collective phases~\cite{Carusotto2009,Boite2013}, interaction-induced hopping \cite{Rodriguez2016}, phase-controlled bistability~\cite{Goblot2016}, cellular automata~\cite{Li2016}, and topological behaviour~\cite{Kartashov2017}.

To model the stochastic choice of the system at the bifurcation point we add a noise term $\theta_n$ to the right-hand side of Eq.~\eqref{eq:GP}, where $\langle\theta_n^*\theta_m\rangle=2\theta_0^2\delta_{n,m}dt$ and $\langle\theta_n\theta_m\rangle=0$. The magnitude of such a term can be controlled experimentally and it can cause jumps between bistable states~\cite{Abbaspour2014}. The probability of a jump is high near the bifurcation point and decreases as one moves further and further beyond the bifurcation point. The choice of bistable state after passing the bifurcation point also depends on the angle of the line along which the parameters are ramped up in the $F_n$-$\Delta_n$ plane. We define an effective classical spin as $s=\pm 1$ depending on whether the system chooses the higher or lower intensity state when in the bistable zone. Fig.~\ref{fig:bifurcation}b shows how the average of this effective spin, obtained by calculating over different realizations of the stochastic noise, varies with $F_\mathrm{init}$. For a value of $F_\mathrm{init}^{0} = 0.1135$, for which the line in Fig.~\ref{fig:bifurcation}a passes between the lower and upper boundaries of the bifurcation zone, we find an equal chance to form high or low intensity states, corresponding to $\langle s\rangle=0$.

To address complex optimization problems, we consider an all-to-all type coupling $J_{nm}$. In principle, this could be realized with a feedback approach~\cite{Byrnes2013} where the optical output of all modes is extracted and fed back into the system after some manipulation. This can be done efficiently using an optical matrix multiplier comprised of a pair of lenses and a spatial light modulator~\cite{McAulayBook,KarimBook,GoodmanBook} (see also \cite{SM}, Sec.~A). An alternative scheme to realize a highly-connected graph can be realized with a bus-coupled mechanism (\cite{SM}, Sec.~B), suitable for the graph partitioning problem.

\begin{figure}[b!]
\includegraphics[width=\columnwidth]{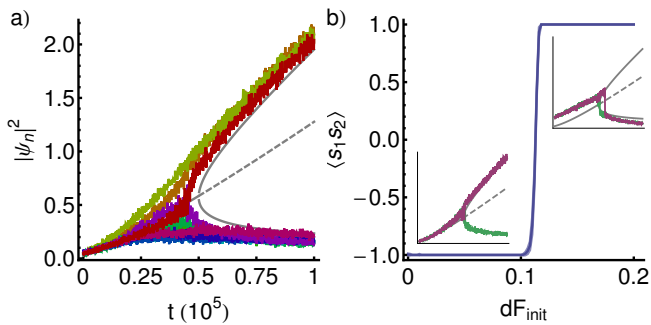}
\caption{a) Colored curves show the evolution of $|\psi_n|^2$ for a ten mode system slowly ramped through the bifurcation point. The solid gray curves show the stable stationary solutions followed by the single mode system in the adiabatic limit; the dashed gray curve shows an unstable branch associated to the bistable region. $J_{nm}$ was chosen as a real symmetric matrix with Gaussian distributed values of root mean square size $0.04$. Other parameters were the same as in Fig.~\ref{fig:bifurcation}, using the unbiased value $F_\mathrm{init}^{0}=0.1135$ corresponding to the vertical dashed line in Fig.~\ref{fig:bifurcation}b. b) Dependence of the effective spin correlation, $\langle s_1 s_2\rangle$, for two antiferromagnetically coupled modes ($J_{12} = 0.04$) on the difference in the initial pump amplitude from the unbiased value. The insets show the time dependence, using the same axes as in a).}
\label{fig:multimode}
\end{figure}

A typical example of the system dynamics for a randomly chosen coupling matrix $J_{nm}$ is shown in Fig.~\ref{fig:multimode}a for $N=10$.
The coherent drive amplitude $F_n(t)$ and detuning $\Delta_n(t)$ are chosen the same for all modes and ramped slowly through the bifurcation point (following the solid line in Fig.~\ref{fig:bifurcation}a). The overall scale of the couplings is taken such that their root mean squared value $\langle J_{nm} \rangle$ is kept small ($\lesssim 0.1$), and that it can be considered as a perturbation to the single mode dynamics. After crossing the bifurcation point, all modes adopt either a high or low intensity, close to the exact adiabatic solutions for the single-mode case. Remarkably, although our $N$-mode system is multistable with $2^N$ stable states, for the particular noise realization used in Fig.~\ref{fig:multimode}a, we find empirically that the system attains the state minimizing the effective Ising Hamiltonian with arbitrary connectivity, $H_\mathrm{eff}=\sum_{n,m}J_{nm} s_n s_m$, corresponding to a spin glass system. An underlying intuition for this behavior can be based on the behavior of the basins of attraction of the system, where we find that two coupled modes with $J_{nm}>0$ are more likely to form in an antiferromagnetic state (see \cite{SM}, Sec. C). Considering the $N=10$ system with different noise realizations, the optimal solution appeared in over $30\%$ of tries. We also found that the success probability increased for slower increase of $F_n(t)$ and $\Delta_n(t)$ (\cite{SM}, Sec. D). Given that polariton lifetime can be around a picosecond, and ultrafast switching of bistable states is well established \cite{DeGiorgi2012,Cerna2013,Cancellieri_PRL2014}, we conservatively predict device operation times of nanoseconds with little loss of success probability. The obtained state is sensitive to $F_\mathrm{n,init}$, as expected from its influence on the single mode behavior (Fig.~\ref{fig:bifurcation}b). Fig.~\ref{fig:multimode}b shows that antiferromagnetically coupled modes may be forced into a ferromagnetic state under sufficient adjustment of $F_\mathrm{init}=F_\mathrm{init}^{0}+dF_\mathrm{init}$.

To characterize the polariton optimizer, we repeated our calculations with different randomly generated coupling matrices $J_{nm}$. The performance was benchmarked by comparison to the ideal ground state of $H_\mathrm{eff}$ using several metrics. The Hamming distance $h_\mathrm{dist}=(1/N)\sum_n^N(1-s_n s^{(g)}_n)/2$ is a measure of how far in configurational space the obtained ($s_n$) and ideal states $s^{(g)}_n$ are. We also defined the energy difference of the approximate solution $E$ and the ground state as $dE=(E-E_g)/(E_\mathrm{max}-E_g)$, where the normalization factor allows to consider various instances on equal footing. Finally, the success probability $p_{\mathrm{success}}$ was defined as the probability of having {\it exactly} the ground state energy.
\begin{figure}[b!]
\includegraphics[width=\columnwidth]{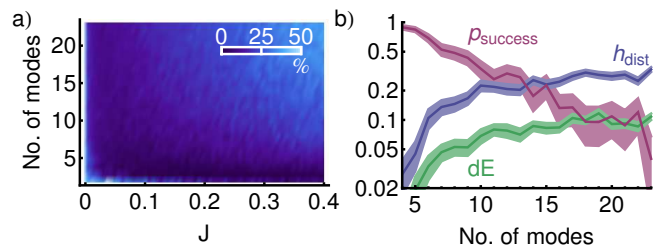}
\caption{a) Variation of the energy difference from the ground state $dE$ with $\langle J \rangle$ and the number of modes in the system. b) Variation of the average Hamming distance $h_\mathrm{dist}$, energy difference from the ground state $dE$, and success probability $p_\mathrm{success}$ (all in logarithmic scale) with the number of modes in the system. Here the value of $\langle J \rangle$ was taken as the optimum from (a) for each number of modes. Parameters were taken the same as in Figs.~\ref{fig:bifurcation} and~\ref{fig:multimode}a, with $2\tau = 10^4$. The shading around the curves indicates the standard error.}
\label{fig:scaling}
\end{figure}

Considering, for simplicity, the zero bias case (setting $F_\mathrm{init} = F_\mathrm{init}^{0}$ as for the single mode system), Fig.~\ref{fig:scaling} shows the variation of the different performance characteristics as the system size is increased. For each system size there is an optimum of the overall scale of the coupling $\langle J \rangle$. While the success probability is less than unity, it is finite, and the ability of an optical system to reach a state in nanoseconds makes it feasible to rerun the simulator several times. The time taken for the simulator to obtain the correct result after several trials is then inversely proportional to the success probability, and scales exponentially with the system size, as is expected for a non-polynomial problem. Even when a state different to the ground state is found, the Hamming distance and energy difference suggest that it is still a reasonable approximation of the ground state.
\begin{figure}
\includegraphics[width=1.\columnwidth]{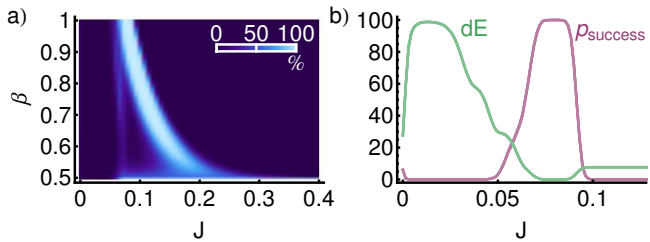}
\caption{a) Variation of the success probability for solving the graph partitioning problem with $J$ and $\beta$ for a graph with $N=6$. b) Variation of the energy difference from the ground state and success probability with $J$ for $\beta=1$. Other parameters were taken the same as in Figs.~\ref{fig:bifurcation} and~\ref{fig:multimode}a, with $2\tau=10^4$. The standard error is within the line thickness.}
\label{fig:graphPolariton}
\end{figure}

\textit{Graph partitioning.} We now show an Ising-type embedding of the NP-hard graph partitioning problem, and solve it emulating the bistability-based polaritonic optimizer. Given the graph $G=(V,E)$ with vertices $V = \{v_j\}_{j=1}^{N_v}$ and the set of edges, $E = \{e_k \}_{k=1}^{N_e}$, the task is to find the partitioning into two groups of equal numbers of vertices (assume $N_v$ being even) such that the number of edges between groups is minimized \cite{Lucas_FroPhy2014}. Such a task can be used to speed up classical calculations with parallelization and can be formulated as the minimization of an all-to-all connected Ising Hamiltonian
%
\begin{equation}
\label{eq:H_graph}
H = J \sum_{(k,l) \in E} (1 - s_k s_l)/2 + J \beta \sum_{(i,j) \in V}^{N_v} s_i s_j,
\end{equation}
where $J$ and $\beta$ are real positive parameters. Here, the first term assures that each connection between the two groups of spin $\pm 1$ introduces an energy penalty, and for $J > 0$ the number of edges will be minimized. The second term represents a constraint, and ensures that the total spin is zero, thus giving equal partitioning for large $\beta$. Rearranging terms in the Hamiltonian \eqref{eq:H_graph} shows that to solve the graph-partitioning problem one shall find the ground state of the Ising network with couplings of two magnitudes, $\beta J$ and $(\beta-1/2)J$.  This for instance can be encoded in the real-space coupled polariton nodes through the common bus (see \cite{SM}, Sec.~B).

As a test, we chose a particular graph partitioning problem with $N=6$ (\cite{SM}, Sec. E) with two degenerate energy configurations (in addition to spin degeneracy), which can make optimization more challenging. Fig.~\ref{fig:graphPolariton} shows that while the overall scale of $J$ and the value of $\beta$ should be carefully chosen, the system can solve the graph partitioning problem. The scaling with the system size for randomly chosen graphs is shown in \cite{SM}, Sec. E.

\textit{Knapsack problem.} As a further example of an NP-hard problem solved by the polaritonic network, we consider the knapsack problem. Having the list of $N$ objects of fixed weight ($w_i$) and cost ($c_i$), we want to fill a knapsack maximizing its cost, given that the maximal total weight is limited to $W_{\mathrm{max}}$. Here, $i$ is an item index running from $1$ to $N$, and we introduce a binary variable $s_i$, which is equal to $1$ when an object is inside the box.
\begin{figure}[h!]
\includegraphics[width=\columnwidth]{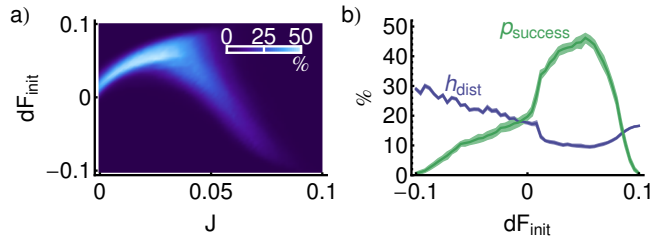}
\caption{a) Success probability for solving the fifteen-mode knapsack problem with a polariton simulator. b) Variation of the success probability and Hamming distance with the overall scale of $dF_\mathrm{init}$, taking optimum values of $J$ for each value of $dF_\mathrm{init}$. Parameters were taken the same as in Fig.~\ref{fig:multimode}, although considering the effective spin Hamiltonian with bias terms representing a specific instance of the knapsack problem. The shading around the curves indicates the standard error and $2\tau=10^4$.}
\label{fig:knapsack}
\end{figure}

The knapsack problem can be formulated as the minimization of an Ising Hamiltonian with bias terms, i.e.,
$H = \sum_{n,m} J_{nm} s_n s_m - \sum\limits_n h_n s_n$
(see \cite{SM}, Sec. F for details). Recalling that the average effective spin population depends on $F_\mathrm{init}$ (as shown in Fig.~\ref{fig:bifurcation}b), we identify the difference of $F_\mathrm{init}$ from the value favoring equal spin populations in the single mode system as an effective biasing parameter $h_n$. As a test we choose an instance of the knapsack problem in its bounded version (see \cite{SM}, Sec. F for details). Using the effective bias, the outlined knapsack problem with 15 spins in total is modelled by time evolving Eq.~\eqref{eq:GP} with the corresponding matrix $J_{nm}$ and the appropriate vector of biases $dF_\mathrm{init}$. The probability of successfully reaching the ground state of the effective Hamiltonian is shown in Fig.~\ref{fig:knapsack} as a function of the overall scale of $dF_n$, where we find a success probability $\sim 50\%$ for optimally chosen $dF_\mathrm{init}$.

\textit{Conclusions.} We have considered nonlinear, resonantly excited, polaritonic-based lattices as coherent driven-dissipative machines which are capable of solving Ising-type optimization problems. We presented an encoding scheme for binary information based on bistable behavior of each mode, which naturally appears for polaritons due to Kerr-type interaction. Two possible schemes for experimental implementation of an all-to-all connected real space polariton network were proposed, and were exploited to solve the graph partitioning and the knapsack problem.  While the results do not suggest improved scaling with system size, showing approximately exponential reduction of the probability of getting the correct ground state, the devices can speed up calculation due to fast operation (few nanoseconds for $N=20$ modes with $\sim 10^6$ configurations) and use of now well-established fabrication techniques of photonic systems. Very recently, concepts of using exciton-polaritons to represent the Potts model have appeared \cite{Kalinin2018A,Kalinin2018B}. As mechanisms of resonantly driven multistability are established experimentally~\cite{Paraiso2010,Cerna2013,Grosso2014} generalization to such models seems within reach, together with consideration of nonresonantly driven mechanisms of bistability~\cite{Klaas2017,Pickup2018}.

\begin{acknowledgments}
We thank N. G. Berloff for useful comments on the manuscript. HS acknowledges support by the Research Fund of the University of Iceland, The Icelandic Research Fund, Grant No. 163082-051. TL was supported by the Ministry of Education (Singapore) grant 2017-T2-1-001.
\end{acknowledgments}


\clearpage

\setcounter{equation}{0}
\setcounter{figure}{0}
\renewcommand{\theequation}{S\arabic{equation}}
\renewcommand{\thefigure}{S\arabic{figure}}

\setcounter{MaxMatrixCols}{10}

\begin{center}
\textbf{Supplemental Material: Solving NP-hard problems with bistable polaritonic networks}
\end{center}

In this Supplemental Material we first present the details of proposed setups that can potentially enable high connectivity for polaritonic graphs (sections A and B). In section C, we provide an analysis of fixed points and their basins of attraction for two coupled modes, to show how effective antiferromagnetic coupling is favoured in our system. Section D shows the dependence of the success probability of solving the Ising problem considered in the main text on the system integration time. Section E gives further details on the graph partitioning problem and its dependence on the system size. Finally, section F gives further details of the solution of the knapsack problem considered in the main text.

\subsection{A: Feedback scheme} \label{SA}

\begin{figure}[h!]
\includegraphics[width=1.\columnwidth]{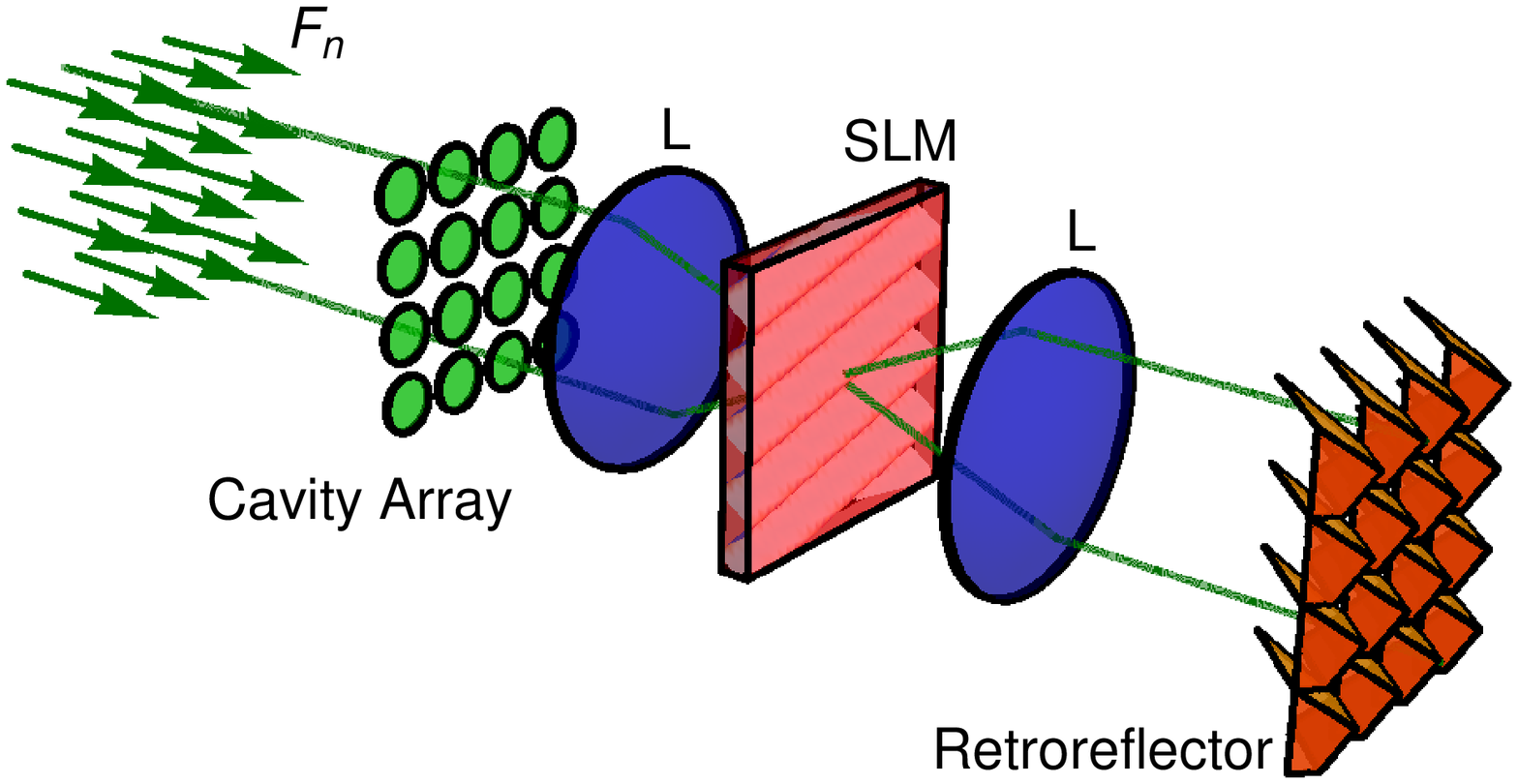}
\caption{Example scheme of all-to-all coupling using an SLM. The cavity array, corresponding to an array of micropillar cavities, photonic crystal cavities, or other set of nonlinear modes, is coherently excited by the driving field, $F_n$. The transmitted light from the cavity array is passed through a Fourier lens (L), which maps the spatially separated modes to Fourier space. If the spatial light modulator (SLM) contains a component at the difference wavevector of the modes, it can allow their coupling in Fourier space via diffraction. The light is retroreflected and returned to the real space after passing back through the original Fourier lens so as to feed back into the cavity array.}
\label{fig:S2}
\end{figure}

We consider a planar polaritonic system, for example where a cavity is formed by distributed Bragg reflectors and the active medium corresponds to a stack of quantum wells, hosting excitonic quasiparticles. The lattice of localized modes can be created by an optical potential, and each mode is fully separated in space and can be driven resonantly. Assuming that the substrate on which the cavity array is grown is transparent, light is emitted through the back surface. When this emitted light passes through a Fourier lens, the localized modes in the real space of the microcavity are mapped to the reciprocal space of a plane behind the lens. There one can place a spatial light modulator (SLM), represented by material with spatially modulated refractive index (Fig.~\ref{fig:S2}). Thus, in this plane we expect a coupling of the different modes according to the weights given by the Fourier components of the effective potential (refractive index variation) set by the spatial light modulator. Placing another mirror or retroreflector behind the spatial light modulator ensures that the light is reflected and fed back into the cavity array. Using the Fourier lens the modes in the reciprocal space of the spatial modulator $\psi_{\mathbf{k}}$ are then mapped back to the real space modes of the active medium, $\psi_{\mathbf{x}}$.

\begin{figure}[t!]
\includegraphics[width=1.\columnwidth]{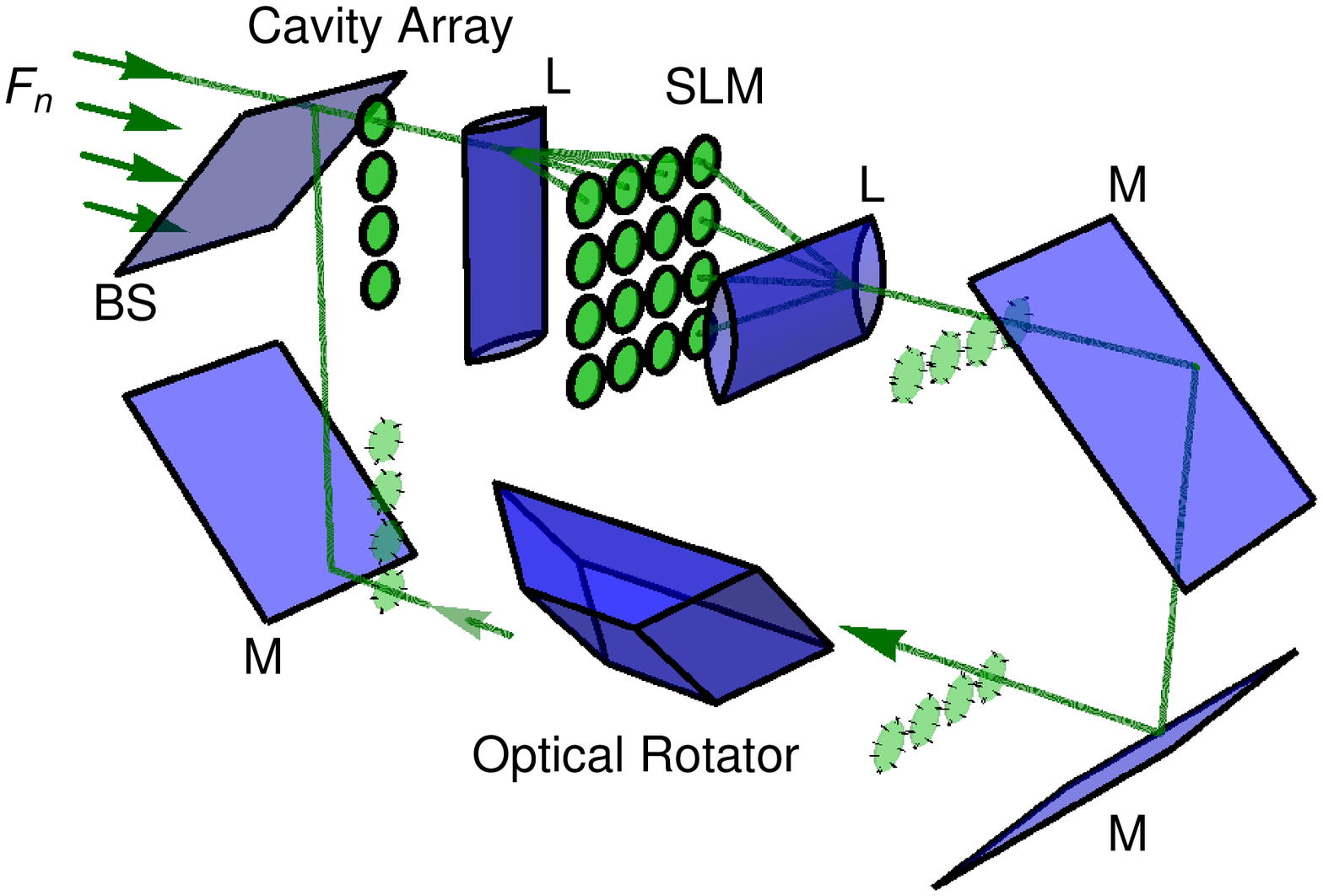}
\caption{Example scheme of feedback loop providing all-to-all coupling. The transmitted light from the cavity array is fed through an optical matrix multiplier, comprised of a combination of lenses (L) and a SLM. The light is then reflected by a series of mirrors (M) and an optical rotator (e.g., Dove prism) before being overlapped with the original driving field, with the aid of a beam splitter (BS), and being fed back into the cavity array.}
\label{fig:S1}
\end{figure}

As an alternative, the traditional optical matrix multiplier method could be employed (Fig.~\ref{fig:S1}). In this case the cavities are arranged into a 1D array. A lens maps their output into an array, which is focused and modulated on a spatial light modulator. The transmitted signal is mapped by a second lens into a 1D array that is returned to the original cavity array via a feedback loop.

As compared to the scheme in Fig.~\ref{fig:S2}, this alternative allows to operate with a lower resolution SLM (e.g., a shadow mask could also be used), while the overall system size is larger since the cavity array is stretched out into a 1D rather than 2D array.

\subsection{B: Bus-coupled scheme}

\begin{figure}
\includegraphics[width=1.\columnwidth]{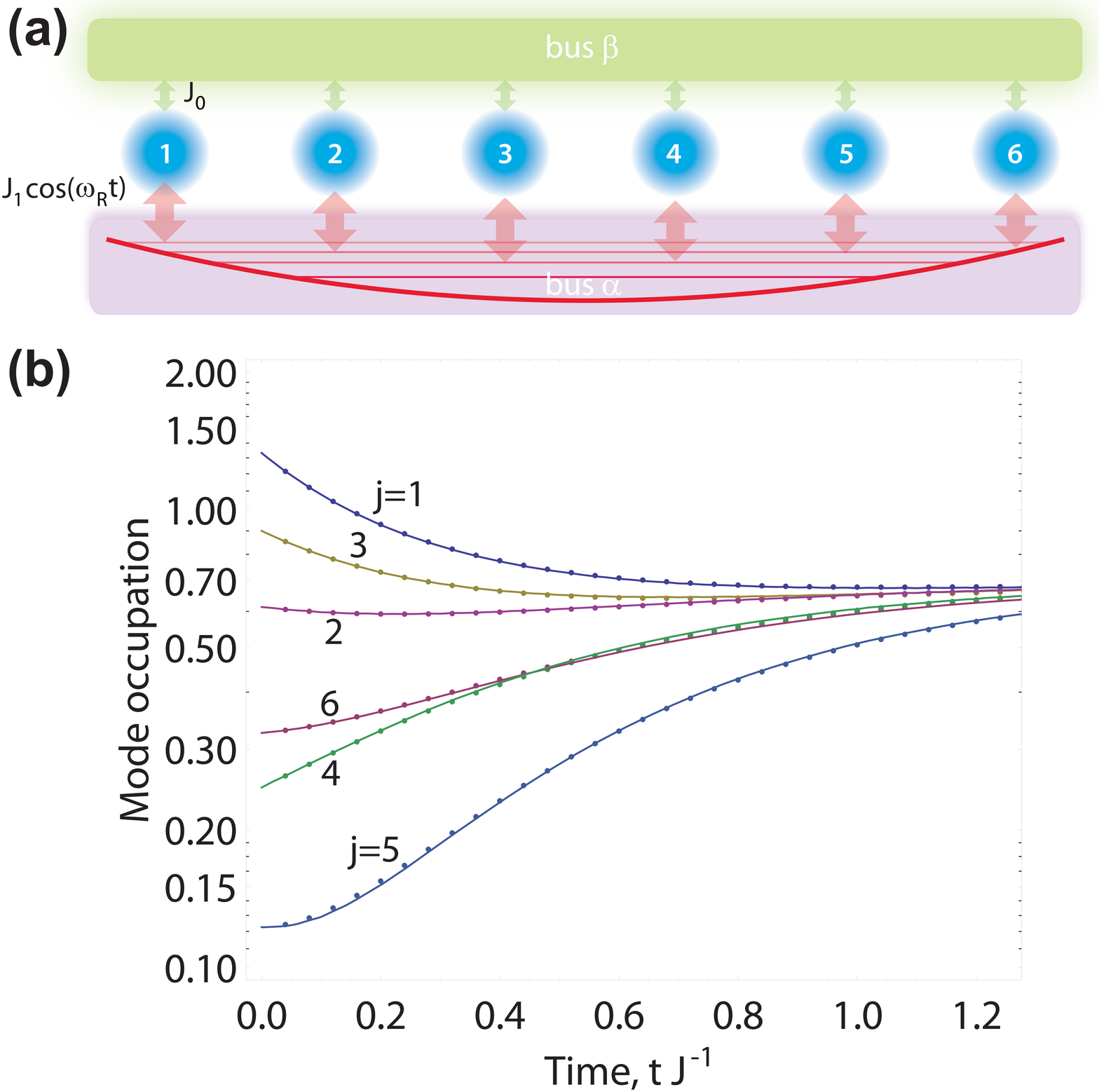}
\caption{(a) Bus-coupled polaritonic boxes. The all-to-all coupling is realized though a detuned common bus (top channel). Additionally, the edge coupling for the particular graph is realized with a multimode bus (bus $\alpha$, bottom) with multiplexed coupling. (b) The proof-of-principle calculation of a six mode system coupled through three bus modes. The plot shows the intensity for each mode ($j=1,..,6$) for the full time dependent calculation (curves) and effective eliminated version (dots), which coincide.}
\label{fig:bus}
\end{figure}

We consider a chain of unconnected polariton wells, where the wavefunction in each box is described by the mean-field $\psi_n$. We take the situation where there is no direct coupling between different polariton boxes, but there is a channel that runs along side the chain which represents a coherent polariton bus, similar to a quantum connecting bus used for microwave circuits~\cite{Song2017}. In particular we consider two types of buses, being geometrically suitable for the graph partitioning problem (bus $\alpha$ and bus $\beta$ in Fig.~\ref{fig:bus}a).

The top channel is described by macroscopic wavefunction $\chi_0$, and is detuned to frequency $f_0$. This would allow to effectively realize homogeneous all-to-all coupling, which is for instance required by the second term in Hamiltonian 3 of the main text. The bottom channel contains a set of modes, described by mean-field $\chi_m$ ($m \geq 1$) and bare frequency $f_m$. If we further imagine that the potential between the channel and each polariton box is modulated in time, then we can write the equations of motion:
\begin{align}
\label{eq:psi_n_t}
\frac{\partial \psi_n}{\partial t}  &= (-i\Delta_n - \kappa_n/2 )\psi_n + F_n -i J_{0} \chi_0  \\ \notag &-i \sum_m J_{nm} \cos (\omega_m t) \chi_m - |\psi_n|^2 \psi_n,\\
\frac{\partial \chi_m}{\partial t} & = -i \omega_m \chi_m  -i \sum_n J_{nm} \cos (\omega_m t) \psi_n - \chi_m,
\\
\frac{\partial \chi_0}{\partial t} & = -i \omega_0 \chi_0  -i  J_{0} \sum\limits_{n} \psi_n - \chi_0,
\end{align}
where $F_n$ is a coherent pump amplitude for the $n$-th mode with detuning $\Delta_n$ and decay $\kappa_n$. Here $J_{nm}$ characterizes the coupling strength between the localized mode $\psi_n$ and channel mode $\chi_m$, which is modulated at frequency $\omega_m$. $J_0$ is the homogeneous coupling to the $\beta$ bus. We account for nonlinear losses in the equation for the localized modes, which may undergo condensation (the amount of nonlinear losses has been scaled to unity through the definition of $\psi$ and $\chi$). We also account for losses of the channel modes, by an amount scaled to unity through the timescale. We consider the bus modes to be non-driven. Now, let $\chi_m = \chi '_m e^{-i \omega_m t}$, such that
\begin{align} \notag
\frac{\partial \chi'_m}{ \partial t} &= -i \sum_n J_{nm} \cos(\omega_m t ) e^{i \omega_m t} \psi_n - \chi'_m \\ &\approx \frac{-i}{2} \sum_n J_{nm} \psi_n - \chi'_m,
\end{align}
where we neglect fast oscillating components. Solving the last equation for the stationary state and assuming fast dynamics for the bus, we can rewrite the system using effective couplings for the modes. Similarly, the stationary solution for $\chi_0$ (static bus) gives
\begin{align}
\chi_0 = -\frac{J_0}{f_0 - i} \sum_n \psi_n.
\end{align}
Substituting the stationary solutions into equation \eqref{eq:psi_n_t} gives the effective couplings. This ultimately allows to arrange $J$ and $J\beta$ terms for the partitioning problem.

To test the validity of the used approximation, we perform the dynamical simulation of an $N=6$ mode system with three dynamical buses, which are detuned by $25$, $10$, and $-15$ energy units (measured by the channel's decay rate), and are coupled through $J_{nm} = 1$. Other parameters are $F_n = \kappa_n = 1$, $J_0 = 0$, and we take random initial conditions. The results are shown in Fig.~\ref{fig:bus}b, and reveal that the mode intensities can be successfully described by the effective theory, where time-dependent coupling is converted into selective intermode interaction.

\subsection{C: Fixed points and basins of attraction}

While we only claim empirical evidence for a heuristic rather than exact global minimizer, there is an intuition underlying the operation of our system. For two coupled modes, the evolution of the system is described by the equations (neglecting the noise terms):
\begin{align}
i \frac{d\psi_1}{dt}&=(-\Delta - i/2 +|\psi_1|^2)\psi_1 + F + J \psi_2,\\
i \frac{d\psi_2}{dt}&=(-\Delta - i/2 +|\psi_2|^2)\psi_2 + F + J \psi_1.
\end{align}
Without coupling (in the absence of $J$) and above the bifurcation point the possible solutions, $(\psi_1,\psi_2)$, are $(\psi_L,\psi_L)$, $(\psi_L,\psi_U)$, $(\psi_U,\psi_L)$, and $(\psi_U,\psi_U)$, where $\psi_L$ and $\psi_U$ are the lower and upper intensity single mode stationary solutions. We recall that when the system is in a stationary solution it evolves with a real energy (which is zero here as we are working in the frame rotating with the pump).

Considering first the state $(\psi_L,\psi_L)$, we are interested in how its energy changes in the presence of coupling $J$. Setting $\psi_{1,2}=\psi_L \exp(-i \omega_{1,2} t)$, that is, allowing the energy of the stationary state to be changed and become complex (such that it is no longer stationary):
\begin{align}
i \frac{d\psi_1}{dt} &= \omega_1 \psi_L = J \psi_L\\
i \frac{d\psi_2}{dt} &= \omega_2 \psi_L = J \psi_L
\end{align}
Here, we have made the crudest approximation on the right-hand side, assuming that the influence of one mode on another is approximately given by taking the influencing mode as being in the single mode solution. While a very crude approximation, it illustrates the principle: $\omega_1$ and $\omega_2$ do not change much from the single mode stationary values when both modes are in the low intensity state.

The same occurs when considering both modes in the upper intensity state, that is, $(\psi_1,\psi_2)=(\psi_U,\psi_U)$. However, considering the case $(\psi_1,\psi_2)=(\psi_L,\psi_U)$, we obtain
\begin{align}
i \frac{d\psi_1}{dt}&=\omega_1\psi_L = J \psi_U,\\
i \frac{d\psi_2}{dt}&=\omega_2\psi_U = J \psi_L.
\end{align}
Here we find that $\omega_1 = J\psi_U/\psi_L$ and $\omega_2 = J\psi_L/\psi_U$. Above the bifurcation point, it is straightforward to find from the analytic solutions that $\psi_U/\psi_L$ has a negative imaginary part, while $\psi_L/\psi_U$ has a positive imaginary part. Thus, the effect of $J>0$ on $(\psi_L,\psi_U)$ is to make $\psi_1$ drop in intensity and $\psi_2$ grow in intensity. This suggests that the state $(\psi_L,\psi_U)$, that is an antiferromagnetic state is more stable due to the coupling $J$. The argument applies in the same way to the state opposite antiferromagnetic state $(\psi_U,\psi_L)$. We can also note that if $J<0$, the antiferromagnetic state will instead be less stable as the signs of $\omega_1$ and $\omega_2$ will be inverted.

Thus, when the system passes the bifurcation point and is fluctuating in the presence of noise, the picture is that because the antiferromagnetic state (for $J>0$) is more stable it is more likely to be chosen by the system. However, to verify this picture we need to consider the basins of attraction of the system in phase space~\cite{Wang2013}. While it is an educated guess that a deeper basin of attraction is also larger in phase space, we are not aware of any law to be certain before actually calculating it.

\begin{figure}
\includegraphics[height=20cm]{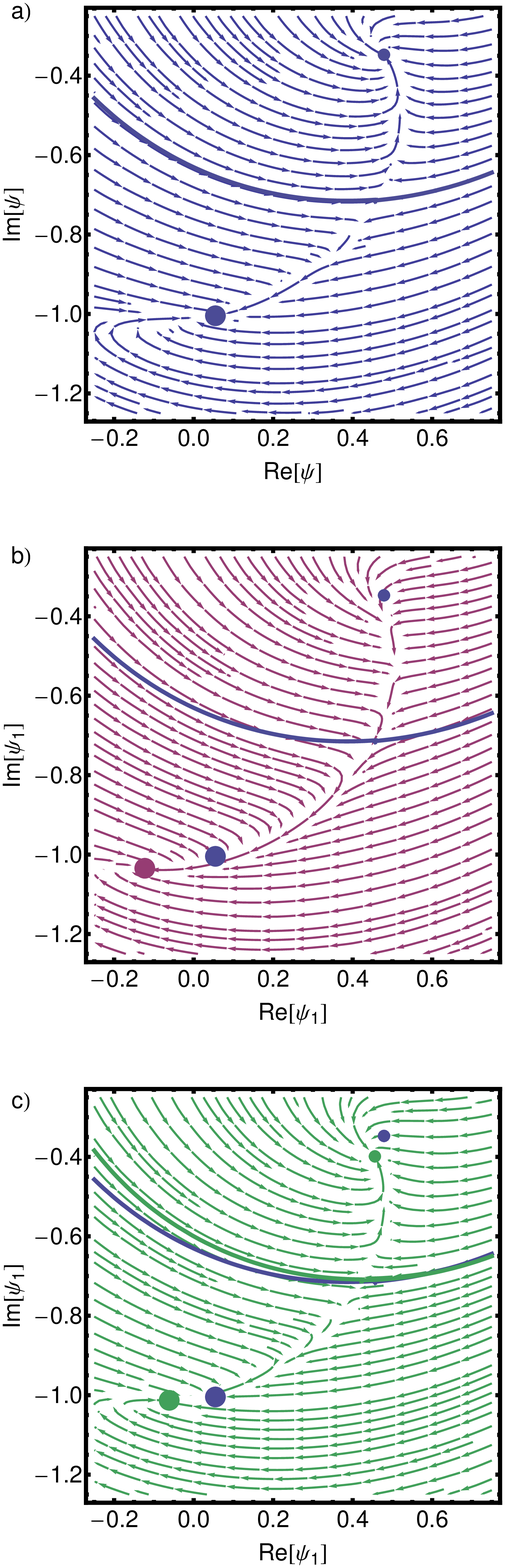}
\caption{Phase diagrams for a single mode system (a, blue), a mode $\psi_1$ coupled to a mode $\psi_2$ in the lower intensity state (b, purple), and a mode $\psi_1$ coupled to a mode $\psi_2$ in the upper intensity state (c, green). Parameters were taken the same as in Fig. 2b, with $dF_\mathrm{init}=0$, $t=1.2\tau$. The large and small spots represent fixed points of high and low intensity, respectively. The solid curves mark separatrices; blue for the single mode system and green for the mode coupled to a mode in the upper intensity state.}
\label{fig:basins}
\end{figure}

Fig.~\ref{fig:basins}a shows the phase diagram of the single mode system. There are two fixed points, corresponding to the low (small dot) and high (large dot) intensity bistable states. The solid curve denotes the separatrix, corresponding to the boundary between basins of attraction between the two fixed points.

Fig.~\ref{fig:basins}b shows how the phase diagram of mode $\psi_1$ is modified by coupling to a second mode $\psi_2$, assuming that the second mode $\psi_2$ is in the lower intensity state (treated with the single mode approximation). Remarkably, not only does the basin of attraction for the upper intensity state grow, as expected from the crude analysis presented above, but it fills the whole phase space as the lower intensity state has become unstable (therefore no separatrix for this case can be plotted and only comparison with case Fig.~\ref{fig:basins}a is shown by the blue solid line). We have also observed the opposite behavior, namely the upper intensity becoming unstable, when $J<0$ (not shown in plot).

Fig.~\ref{fig:basins}c shows the phase diagram of mode $\psi_1$ when coupled to a second mode in the upper intensity state (again treated with the single mode approximation). In this case the low intensity state has stabilized. Although the separatrix is little changed from the single mode case, the shown behavior allows the antiferromagnetic state to be stable in our system.

We note that while the phase diagrams in Fig.~\ref{fig:basins} describe well the mechanism at play for a two coupled mode system, generalization to larger systems is not obvious. As with all heuristic approaches, justification is only possible with empirical testing corresponding to sampling multiple trajectories of the multi-dimensional phase space.


\subsection{D: Dependence on system integration time}

\begin{figure}[t!]
\includegraphics[width=1.\columnwidth]{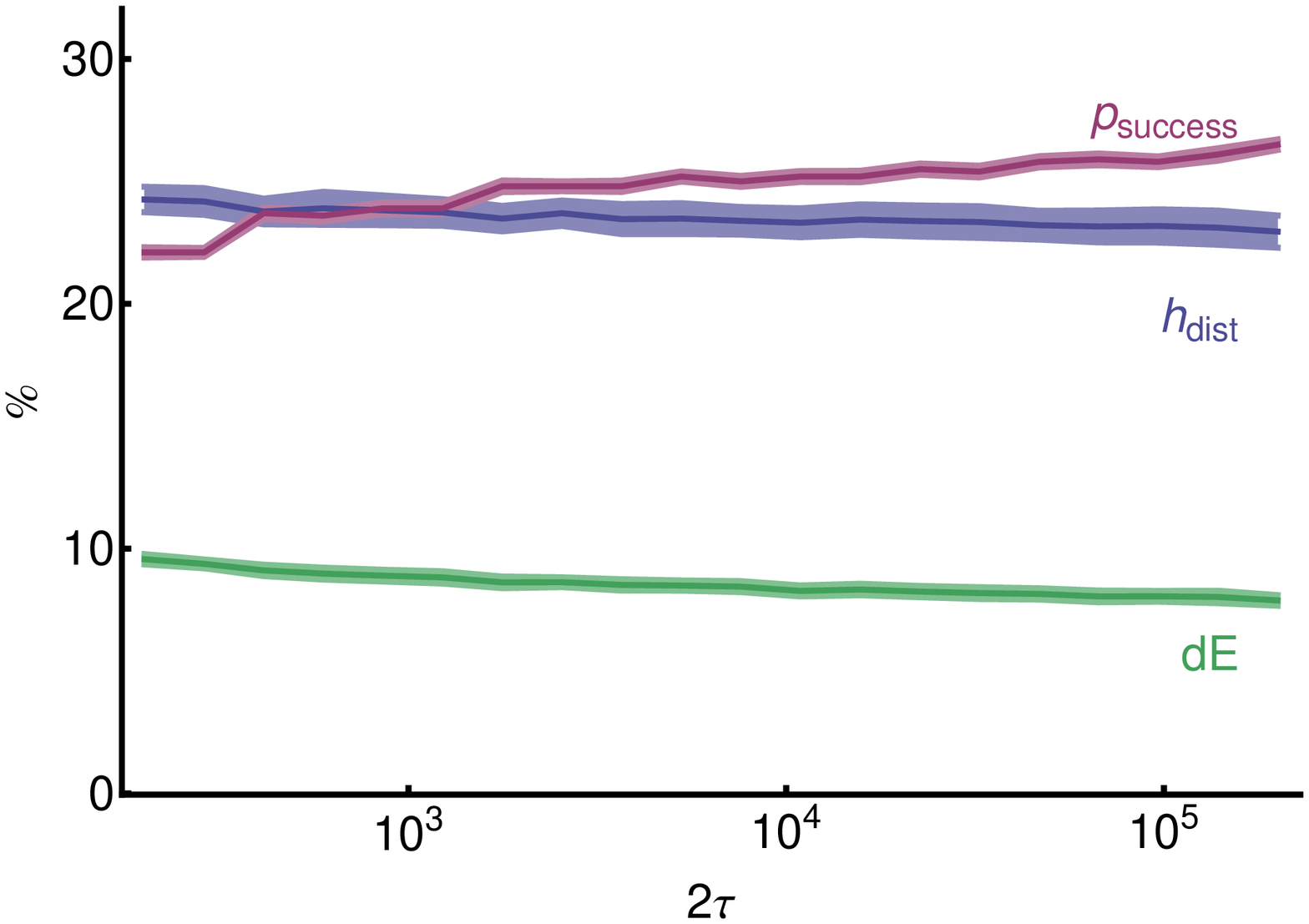}
\caption{Variation of the average Hamming distance, normalized energy difference from the ground state, and success probability as a function of the system integration time $2\tau$. The shading around the curves indicates the standard error. Parameters: $F^0_\mathrm{init}=0.1135$, $J=0.04$, $\theta_0=0.4$.}
\label{fig:Tdependence}
\end{figure}

The performance of the considered polariton simulator depends weakly on the rate at which the parameters are ramped through the bifurcation point. Taking a system of $10$ modes, with randomly chosen couplings and no bias, as in Fig.~3 of the main text, we show in Fig.~\ref{fig:Tdependence} the variation with the system integration time. Here $\tau$ represents the time at which the bifurcation point is reached and $2\tau$ is the total integration time at which the system state is measured. The results suggest overall but weak improvement for slower operation times. However, in practice it may be viable to take smaller $\tau$, while performing more repetitions.

\subsection{E: Details of graph partitioning problem and its scaling with system size}

\begin{figure}
\includegraphics[width=0.8\columnwidth]{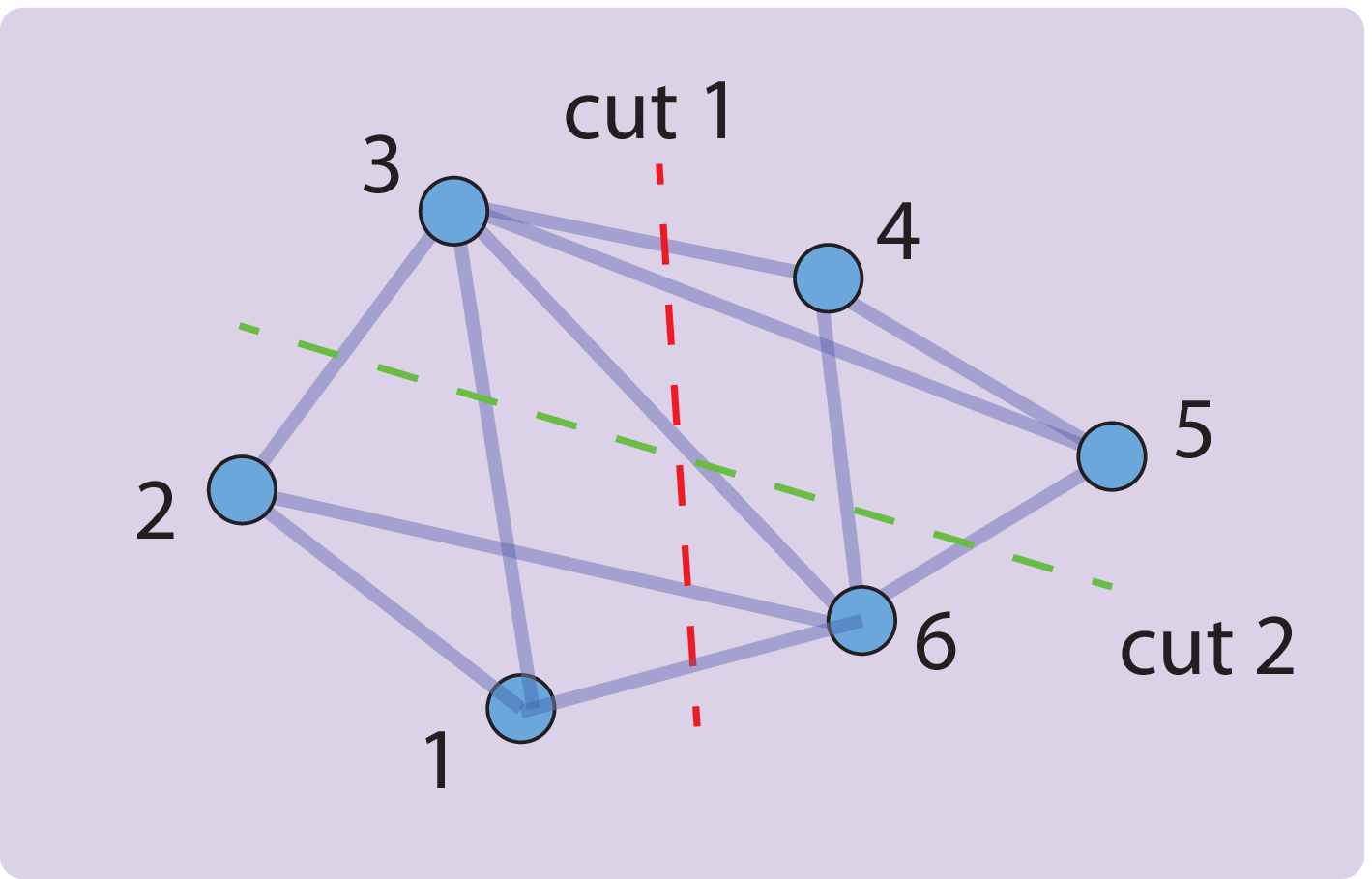}
\caption{Sketch of the graph partitioning problem, aiming to find two separated sets of modes. For the chosen connection net, the ground state solutions are represented by two possible cuts (red and green lines).}
\label{fig:graph}
\end{figure}
\textit{Example.} As a particular test, we can choose the small system of size of $N_v = 6$ vertices, where
\begin{align} \notag
E =  \{&(1, 2), (2, 3), (3, 4), (4, 5), (5, 6), \\
& (1, 3), (3, 5), (4, 6), (3, 6), (2, 6), (1, 6)\}
\end{align}
corresponds to a dense connection grid, where number of connections for each node is larger than the number of nearest neighbors. The problem's topology is depicted in Fig.~\ref{fig:graph}. The optimal partitioning then corresponds to two choices, being $[\{1, 2, 3\}, \{4, 5, 6\}]$ and $[\{1, 2, 6\}, \{3, 4, 5\}]$ (see corresponding cuts 1 and 2 in Fig. \ref{fig:graph}), which are encoded in the $\{-1, -1, -1, 1, 1, 1\}$ and $\{-1, -1, 1, 1, 1, -1\}$ spin configurations (or their bit-flipped partners).

\begin{figure}
\includegraphics[width=1.\columnwidth]{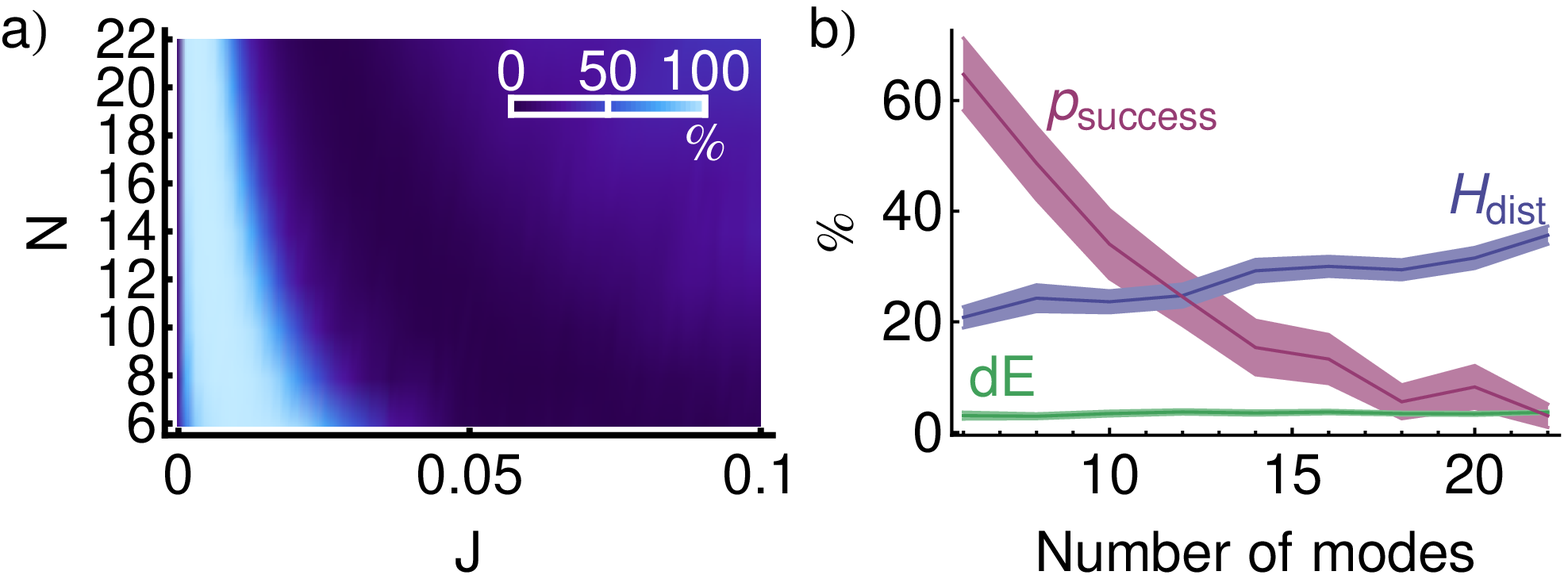}
\caption{a) Variation of the energy difference $dE$ from the true ground state with the system size and overall scale of $J$ for the graph partitioning problem. b) Variation of the energy difference $dE$, success probability, and Hamming distance with the system size. Here we take the optimum value of $J$ for each number of modes. Other parameters were taken the same as in Fig.~4 of the main text.}
\label{fig:graphScaling}
\end{figure}

\textit{Scaling.} To consider the scaling of the graph partitioning problem considered in Fig.~4 of the main text, we average over randomly selected graphs. Each graph is taken to have a number of connections equal to one half the total number of possible connections for given system size (similar results can be obtained for different connection fractions). The results are shown in Fig.~\ref{fig:graphScaling}.

\subsection{F: Knapsack problem details and example}

\begin{figure}[t!]
\includegraphics[width=1.\columnwidth]{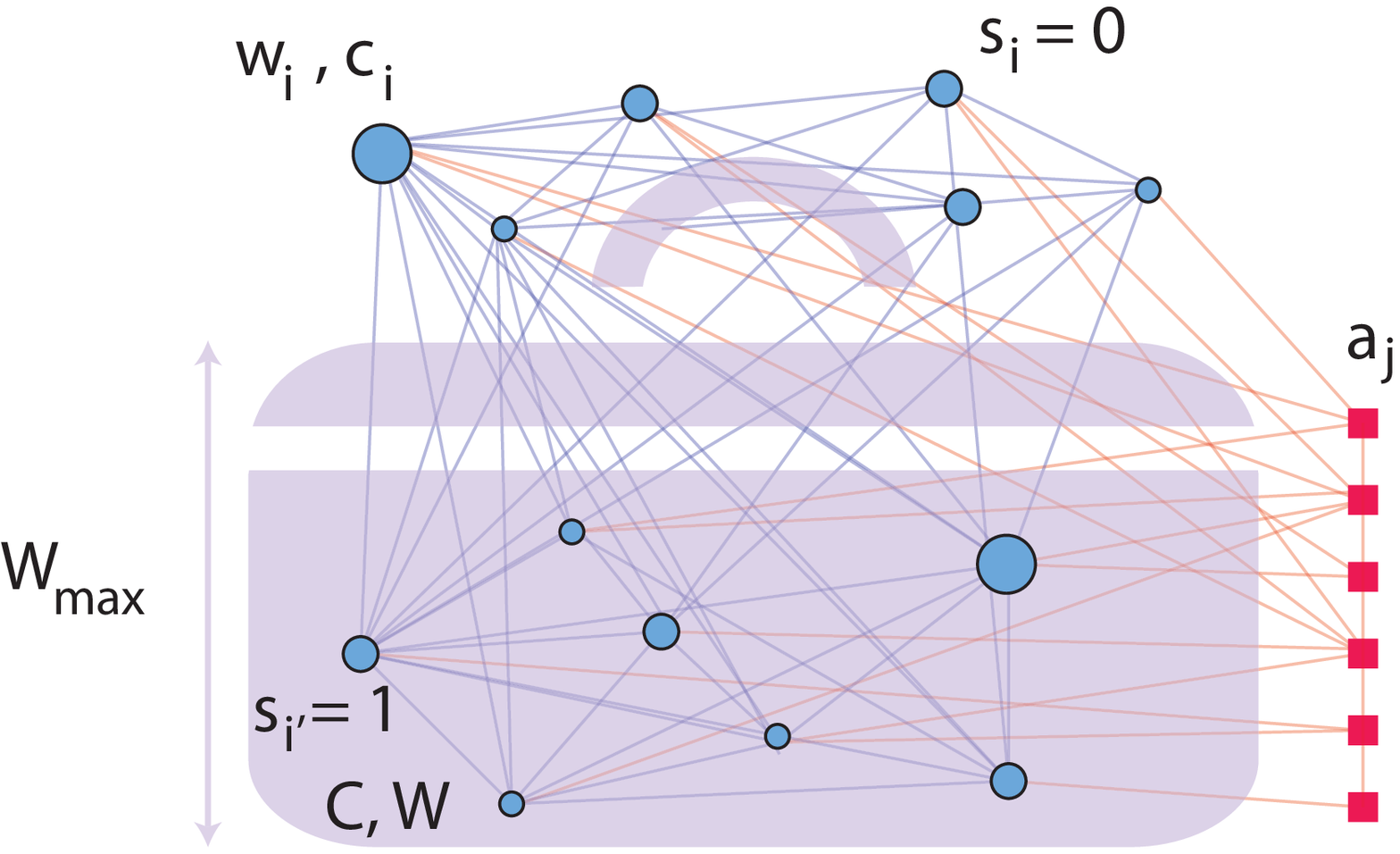}
\caption{Sketch of the knapsack problem. Different items of weight $w_i$ and cost $c_i$ can be placed in the suitcase (assign binary variable $s=1$), or left outside ($s=0$). The maximal weight of the suitcase is bounded by $W_{\mathrm{max}}$. The solution can be obtained by searching for minimal energy configuration for combined item ($s_i$) and auxiliary ($a_i$) spins, combined into an all-to-all connected Ising network.}
\label{fig:KnapsackSketch}
\end{figure}

To formulate the knapsack problem it is convenient to begin with binary variables in the $(0,1)$ basis first. The spin variables $s_i$, with $i$ running from $1$ to $N$ are equal to $1$ when an object is inside the knapsack and $0$ otherwise (see Fig.~\ref{fig:KnapsackSketch}).
The total weight and total value then read $W = \sum_{i=1}^{N} w_i s_i$ and $C = \sum_{i=1}^{N} c_i s_i$, respectively. We further introduce auxiliary binary variables $a_j$, where the index $j$ runs from $1$ to $W_\mathrm{max}$, being the maximal weight. The classical Hamiltonian corresponding to the problem then reads
\begin{align}
\label{eq:H}
H = &\alpha \Big( 1 - \sum\limits_{j=1}^{W_{\mathrm{max}}} a_j \Big)^2 + \alpha \Big( \sum\limits_{j=1}^{W_{\mathrm{max}}} j a_j - \sum\limits_{i=1}^{N} w_i s_i \Big)^2 \\ \notag &- \beta \sum\limits_{i=1}^N c_i s_i,
\end{align}
where $\alpha$ and $\beta$ are parameters for the simulation, chosen such that the global minimum of (\ref{eq:H}) corresponds to the solution. The Hamiltonian (\ref{eq:H}) can be recast as the standard all-to-all connected Ising model with bias terms $h_n$ as
\begin{equation}
\label{eq:H_Ising}
H = -\sum\limits_{n<m}^{N+W_{\mathrm{max}}} J_{nm} s_n s_m - \sum\limits_{n=1}^{N+W_{\mathrm{max}}} h_n s_n + \mathrm{const},
\end{equation}
where $J_{nm}$ denotes Ising coupling matrix, formed by weights, and $h_n$ is an effective magnetic field formed by the combination of cost and weight.

As a particular instance of the bounded version of the knapsack problem, we consider the example with 3 coins of weight 1 and value 5 (coin $a$), 2 coins of weight 2 and value 10 (coin $b$), and 1 coin of weight 3 and value 25 (coin $c$). The maximal weight is fixed to $W_\mathrm{max} = 9$. Using a brute force algorithm, which considers all possible item (i.e. spin) configurations, the solution is 2 $a$-coins, 2 $b$-coins, and 1 $c$-coin. In the classical spin language these are three degenerate configurations $\{ s \} = \Big[ \{ 0, 1, 1, 1, 1, 1 \}, \{ 1, 0, 1, 1, 1, 1 \}, \{ 1, 1, 0, 1, 1, 1 \} \Big]$.


\end{document}